\documentclass[aps,prl,twocolumn,groupedaddress,amsmath,amssymb]{revtex4}
\usepackage{subfigure}
\usepackage{amsfonts}
\usepackage{amsmath}
\usepackage{amssymb}
\usepackage{subfigure}
\usepackage{soul}
\usepackage{xcolor,soul}
\usepackage[utf8]{inputenc}
\usepackage{amssymb,amsthm}
\usepackage{amsmath}
\usepackage{graphicx,color}
\usepackage{bm}

\newcommand{\im}[0]{\imath}

\begin{document}

\title{Extreme Non-Reciprocal Near-Field Thermal Radiation via Floquet Photonics}

\author{Lucas J. Fern\'andez-Alc\'azar$^{1}$, Huanan Li$^{2}$, Tsampikos Kottos$^1$} 
\affiliation{$^1$Wave Transport in Complex Systems Lab, Department of Physics, Wesleyan University, 
Middletown, CT-06459, USA\\
$^2$Photonics Initiative, Advanced Science Research Center, CUNY, NY 10031, USA}
\date{\today}

\begin{abstract}
By utilizing Floquet driving protocols and interlacing them with a judicious reservoir emission engineering
we achieve extreme non-reciprocal thermal radiation. We show that the latter is rooted in an interplay between
a direct radiation process occurring due to temperature bias between two thermal baths and the modulation 
process which is responsible for pumped radiation heat. Our theoretical results are confirmed via time-domain
simulations with RF circuits.
\end{abstract}

% \pacs{Valid PACS appear here}
% \keywords{Suggested keywords}

\maketitle

{\it Introduction -} Thermal radiation is associated with the conversion of the thermal motion of (quasi-)particles, 
in matter with some finite temperature, into electromagnetic emission. Its management constitutes a major 
challenge with both fundamental and technological ramifications \cite{VP07,HSM10,BLID11,WYMWD17,F17,
LF18,CV18,BXNKAK19}. For example, some of the ongoing investigations aim to establish paradigms that 
challenge fundamental limitations in thermal radiation, set by Kirchhoff's emissivity-absorptivity equivalence 
law \cite{K60,ZF14,HSA16,MZF17,GBBM18} and by Planck's upper bound of thermal emission \cite{BA16,
MST16,FFFVC18a,FFFVC18b}. In parallel, other studies exploit the applicability of recent proposals for radiation 
control to daytime passive radiative cooling \cite{RRF13,RAZRF14,GS15,KJCFM17,ZMDZLTYY17}, radiative 
cooling of solar cells \cite{ZRWAF14,ZRF15,LSCZF17}, energy harvesting \cite{RF09,B10,G12,L15,ZSSB16,
B16,F18}, thermal camouflage \cite{LBYLQ18,K14}, etc. It turns out that the implementation of subwavelength 
photonic circuits reinforces the importance of evanescent waves in radiation and allows us to bypass 
the constraints set by Kirchoff's and Planck's laws. This symbiosis of nanophotonics and thermal radiation led 
to the establishment of thermal photonics, which holds promises for novel technologies in energy harvesting 
and near-field thermal radiation management \cite{RKT89,CW51}.

A long-standing problem in thermal radiation management is the quest for novel non-reciprocal devices, like 
thermal diodes and circulators, that control the directionality of photon emissivity. Along these lines, researchers 
have proposed a variety of schemes ranging from magneto-optical effects \cite{A16,OMAB19a,OMAB19b} to 
non-linearities \cite{AB13,INIT14,FTZMBBBBAMR18,KZR15} and active photonic circuits \cite{LAESK19,BLF20} for 
enforcing directional thermal radiation.

\begin{figure}
\begin{center}
\includegraphics[width=0.45\textwidth]{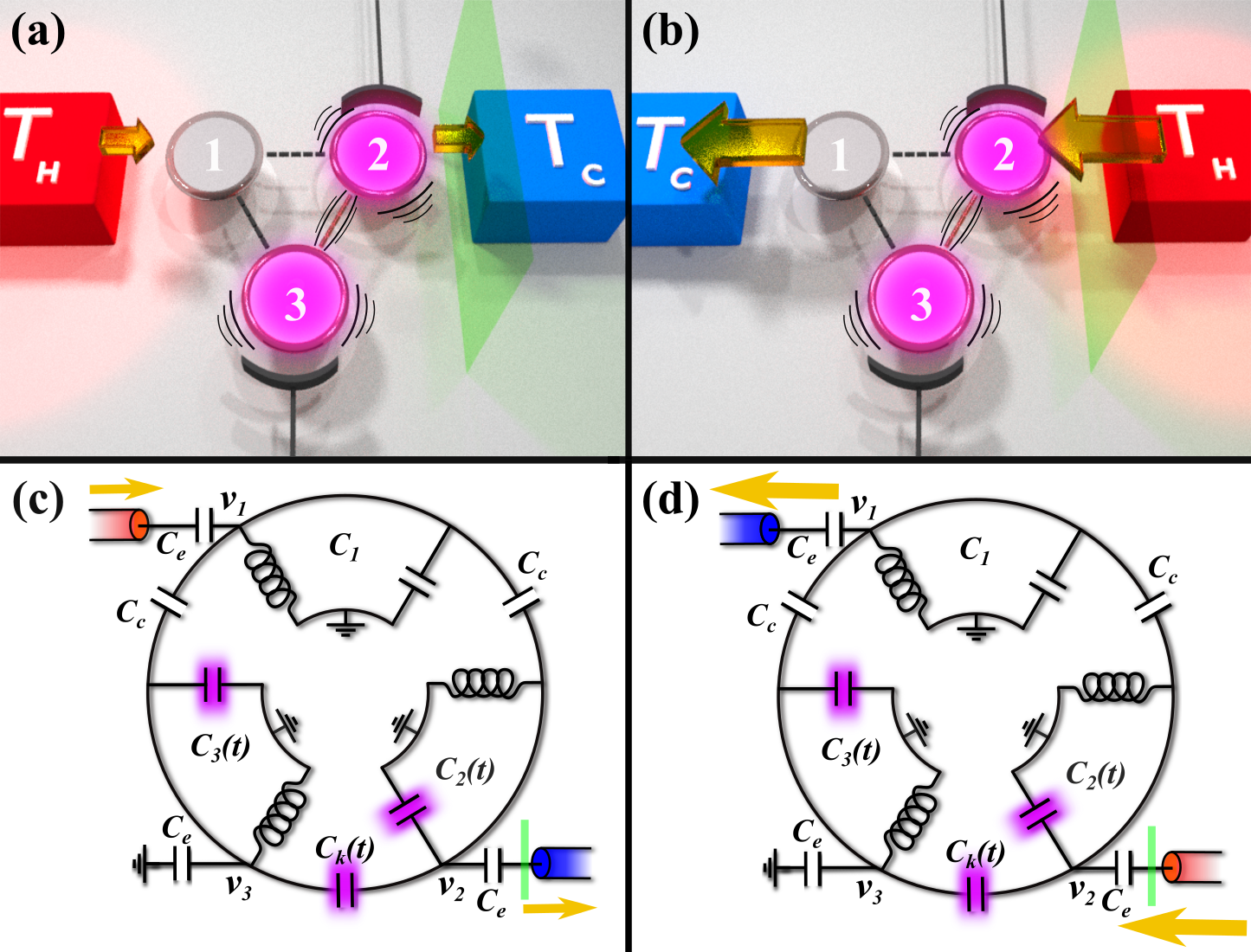}
\caption{(Upper raw) A photonic Floquet diode for near-field thermal radiation. The resonators $n=2,3$ and the
 coupling between them are periodically modulated in time while the $n=1$-resonator is static. (a) In the ``forward" 
(f) configuration, the reservoir with the high (low) temperature $T_{\alpha=1}=T_H$ ($T_{\alpha=2}=T_C$) is 
coupled to the $n=1 (n=2)$-resonator. (b) In the ``backward" (b) configuration the low (high) temperature reservoir $T_{\alpha=1}=T_C (T_{\alpha=2}=T_H)$ is coupled to the $n=1 (n=2)$-resonator. In both cases, the radiation 
current, is measured at lead $\alpha=2$ (green transparent plane). (Lower raw) An equivalent electronic circuit, 
consisting of three capacitively coupled LC resonators. The resonators $n=2,3$ and their coupling are driven 
by modulating the (pink) capacitances.  
(c) Forward configuration and (d) Backward configuration. The currents in both cases are measured at the same 
position at the $\alpha=2$ transmission line (bold green line).} 
\label{fig1}
\end{center}
\end{figure}

Here we unveil an interplay between three elements that control the efficiency of thermal rectification in Floquet-driven 
circuits: (a) a judiciously engineered bath emissivity (via photonic filters) of the thermal reservoirs; (b) an 
appropriately designed Floquet protocol that enforces a time modulation of the constituent parameters of a 
photonic circuit; and (c) the temperature gradient between two thermal reservoirs which are coupled resonantly with 
the circuit. The latter is responsible for a biased current while the second element is generating pumped thermal 
radiation which can balance the biased thermal current in one specific direction. We utilize these elements for the 
design of optimal reconfigurable Floquet-based thermal diodes and validate the theoretical predictions via time-domain
simulations. 

%=============================================
{\it Statistical Coupled-Mode-Theory Modeling -- } We consider a photonic network of $N$ coupled modes. The field
dynamics in such a network is described by a time-dependent effective coupled-mode-theory (CMT) Hamiltonian $H_0
(t)=H_0(t+{2\pi\over\Omega})$. We assume that two of these modes are connected directly to two reservoirs 
characterized by temperatures $T_{\alpha=1}\neq T_{\alpha=2}$, see Fig. \ref{fig1}. At thermal equilibrium, the mean 
number of emitted photons at a frequency $\omega$ is $\Theta_{\alpha}(\omega)=\left\{ \exp\left[\hbar\omega/(k_B 
T_\alpha)\right]-1 \right\}^{-1}$. We study the radiative energy transfer between these reservoirs for a forward (Fig. 
\ref{fig1}a) and a backward (Fig. \ref{fig1}b) configuration. The process is modeled by a temporal-CMT which takes the 
form \cite{H00} 
\begin{eqnarray}
-\im \frac{d \left| \psi (t) \right \rangle}{dt}  &=& H_{\rm eff} \left| \psi (t) \right \rangle - \im D^T \left| S^+(t) \right \rangle; 
\  H_{\rm eff}=H_0(t) +\im \Gamma \nonumber\\
\left| S^- \right \rangle &=& -\left| S^+ \right \rangle + D \left| \psi \right \rangle,
\label{CMT}
\end{eqnarray} 
where the amplitudes $\left| \psi \right \rangle =\left(\psi_1,\cdots,\psi_N\right)^T$ are normalized such that 
$|\psi_n|^2$ represents the energy in the $n=1,\cdots,N$-th mode. The matrix $\Gamma_{nm}=\gamma_{\alpha} 
\delta_{n,\alpha} \delta_{nm} + \Sigma_{nm}(t)$ represents the dissipation of the $n$-th mode, where 
$\Sigma_{nm}(t)$ describes driving-induced losses and/or gain and $\gamma_{\alpha}$ is the dissipation due to 
coupling of the $n-$th mode with the reservoir $\alpha$. From the fluctuation-dissipation relation, we also have that 
$D_{n,\alpha} = \sqrt{2\gamma_{\alpha}} \delta_{n,\alpha}$. Finally, the complex fields $S^{\pm}_{\alpha}(t)=\left\langle 
\alpha | S_\pm (t) \right \rangle= \int_{0}^{\infty} S^{\pm}_{\alpha}(\omega) e^{\im \omega t}d \omega$ indicate the 
incoming ($+$) and outgoing ($-$) thermal excitations from and towards the $\alpha$-th reservoir. The amplitudes 
$S_{\alpha}^{+}(\omega)$ satisfy the relation
\begin{equation}
\label{bath}
\left\langle  [S^{+}_{\alpha^\prime}(\omega^\prime)]^* S^{+}_{\alpha}(\omega) \right \rangle ={\hbar\omega\over 
2\pi}\phi_{\alpha}(\omega)\Theta_{\alpha}(\omega)\delta(\omega-\omega')\delta_{\alpha,\alpha^\prime},
\end{equation}
where $\phi_{\alpha}(\omega)$ describes spectral filtering of the $\alpha-$th thermal reservoir. Existing proposals 
for the control of spectral emissivity of the thermal reservoirs include the deposition of photonic crystals that support 
band-gaps, or their coupling to the photonic circuit via a waveguide or a cavity with cut-off frequencies, etc. \cite{F17,
LF18,CV18}. For electronic circuits (Figs. \ref{fig1}c,d) the spectral control of the reservoir can be arranged via 
synthesized noise sources. 

{\it Floquet Scattering for Thermal Radiation--} 
In Floquet scattering, an incident excitation $S^{+}_\alpha (\omega)$ at frequency $\omega$ can change its frequency 
by $\pm l\Omega$ and scatter out of the modulated target at a Floquet channel $\omega_l=\omega+l\Omega$ where 
$l \in (-\infty,\cdots,\infty)$. The Floquet scattering matrix ${\cal S}^F$, connecting the outgoing to the 
incoming field amplitudes $\vec{S}^\pm = [ \cdots, \left| S_\pm (\omega_{+1}) \right \rangle , \left| S_\pm (\omega_0) 
\right \rangle , \left| S_\pm (\omega_{-1}) \right \rangle, \cdots ]^T$, is evaluated using Eq. (\ref{CMT})
\begin{equation}
 {\cal S}^F=-I - \im [D] G^F [D]^T; \quad G^F=\left(\omega I -\hat{H}_Q\right)^{-1},
 \label{Floquet_S}
\end{equation}
where $[D]$ represents a block diagonal matrix with blocks $D$, and $G^F$ is the Green's function associated 
with the Floquet Hamiltonian $\hat{H}_Q$. The latter takes the form $\left\langle l,n\right|\hat{H}_Q \left| l',n' \right
\rangle={\Omega\over2\pi}\int_0^{2\pi\over\Omega} dt e^{-\im(l'-l)\Omega t} H_{\rm eff}(t)
- l\Omega\delta_{l,l'}\delta_{n,n'}$ \cite{GD14,E17,EA15,LKS18}.
Using Eq. (\ref{Floquet_S}) we have calculated the average energy 
current \cite{suppl}
\begin{equation}
\bar{I}_{\alpha} =  \int \frac{d\omega}{2\pi} \sum_\beta {\cal T}_{\alpha,\beta}^F(\omega) \left[\hbar \omega 
\Theta_\beta(\omega)\right],
\label{current}
\end{equation}
where ${\cal T}_{\alpha,\beta}^F(\omega)=\sum_l\left(-\delta_{\alpha,\beta}\delta_{l,0}+\left| {\cal S}^F_{\alpha,\beta}
(\omega_l,\omega) \right|^2\right)$ is the total transmittance of all incoming waves at frequency $\omega$ from 
the $\beta-$th reservoir, which are emitted at frequencies $\omega_l$ at reservoir $\alpha$. A positive value 
of $\bar{I}_{\alpha}$ indicates that current flows toward the $\alpha$-th heat bath. 

Equations (\ref{Floquet_S},\ref{current}) extend the standard treatment of thermal radiation to periodically modulated 
photonic circuits and provide a bridge with the field of Floquet engineering \cite{LKS18,LSK18,LK19}. It turns out that 
time-dependent perturbations could induce non-reciprocal transmittance ${\cal T}_{\alpha,\beta}^F(\omega)\ne{\cal T}_{
\beta,\alpha}^F(\omega)$ \cite{SA17,CATSAL18,WMDWSF20} whose origin is traced to interference effects 
between different paths in the Floquet ladder \cite{LKS18}. At the same time, Eqs. (\ref{Floquet_S},\ref{current}) 
emphasize the fact that while non-reciprocal transmittances ${\cal T}_{\alpha,\beta}^F(\omega)\ne  {\cal T}_{\beta
\alpha}^F(\omega)$ are a necessary condition, they are not sufficient for the establishment of non-reciprocal thermal 
radiation. In fact, the integration over frequencies with a weight $\Theta_{\alpha/\beta}(\omega)$ might 
suppress the existence of non-reciprocal heat flux or even restore reciprocity.

%===================================================================================
{\it Rectification Efficiency -} We consider three single-mode resonators $n=1,2,3$, equally coupled with one another, 
see Figs. \ref{fig1}a,b. The first and the second resonators are at the proximity of two reservoirs with temperatures 
$T_H>T_C$. We compare the 
emitted energy flux ${\bar I}_{\alpha}$ at a reference reservoir (say reservoir $\alpha=2$) for two different configurations: 
(i) The forward (f) configuration where the cavity $n=1$ is in the proximity of the hot reservoir i.e. $T_{\alpha=1}=T_H$ 
and the cavity $n=2$ is coupled to a cold reservoir i.e. $T_{\alpha=2}=T_C<T_H$ (see Fig. \ref{fig1}a ). (ii) The backward 
(b) configuration (see Fig. \ref{fig1}b) where $T_1=T_C< T_H=T_2$. The non-reciprocal efficiency of the circuit is 
described by the rectification parameter $\cal{R}$
\begin{equation}
 {\cal R}\equiv{\bar I _2 ^{(f)} 
- (-\bar I _2 ^{(b)}) \over \bar I _2 ^{(f)} + (-\bar I _2 ^{(b)}) },
 \label{eqR}
\end{equation}
where ${\cal R}=\pm 1$ indicates perfect diode action, while ${\cal R}=0$ corresponds to completely reciprocal 
radiation. A rectification parameter $|{\cal R}|> 1$ indicates that the photonic circuit operates as a ``refrigerator". We 
will assume that $T_C$ is fixed. A qualitative understanding of the effects of a temperature gradient $\Delta T\equiv 
T_H-T_C$, modulation frequency $\Omega$, and spectral filtering $\phi(\omega)$ on ${\cal R}$, 
is achieved by analyzing the slow driving limit $\Omega\rightarrow 0$. 

In the forward configuration, the current Eq. (\ref{current}) is approximated as the sum of two contributions \cite{suppl,NFLK20}
\begin{equation}
   \bar{I}_{2}^{(f)} \approx \bar{I}_{2,b}^{(f)} + \bar{I}_{2,p}^{(f)},
   \label{current2}
\end{equation}
where $\bar{I}_{2,b}^{(f)}$ is the current due to temperature bias and $\bar{I}_{2,p}^{(f)}$ is a pumped current associated 
with the time modulation of the circuit \cite{LAESK19}. Further progress is made by considering the 
classical limit ($\Theta_\beta(\omega)\approx k_B T_\beta/(\hbar \omega)$) where
\begin{eqnarray}
\bar{I}_{2,b}^{(f)}&\approx&{\hat {\cal T }} \  k_B(T_1-T_2), \ \ {\hat {\cal T }}=\int \frac{d\omega}{2\pi} {\bar {\cal T }}(\omega); \label{current3}\\
\bar{I}_{2,p}^{(f)}&\approx& \frac{\Omega}{2\pi} {\hat {\cal P }} \ k_B T_0 ; \nonumber
{\hat {\cal P }} =\int \frac{d\omega}{2\pi}\frac{\im}{\omega} \int_0^{2\pi\over\Omega} dt 
\left( \frac{d S^t}{dt}(S^t)^{\dagger} \right)_{2,2}, \nonumber  
\end{eqnarray}
where $T_0=\frac{T_1+T_2}{2}$ and the averaged (over one modulation cycle) transmittance $\bar{{\cal T}}(\omega)
=\frac{\Omega}{2\pi}\int dt |S^t_{21}(\omega)|^2$ can be evaluated using the instantaneous scattering matrix $S^t$.
Equations (\ref{current3}) valid up to ${\cal O}(\Omega \times \Delta T/T_0)$. Notice that $\bar{I}_{2,p}^{(f)}$ 
is proportional to $\Omega$ but independent of $\Delta T$.

Following the same analysis, we evaluate $\bar{I}_{2}^{(b)}$. It turns 
out that its bias component is $\bar{I}_{2,b}^{(b)}= - \bar{I}_{2,b}^{(f)}$ while the pumping current is $\bar{I}^{(f)}_{2,p}\approx \bar{I}^{(b)}_{2,p}$. It is, therefore, 
possible to find a set of parameters $(\Delta T^*,\Omega^*)$ such that the current in the forward (backward) configuration 
$\bar{I}_{2}^{(f)} \approx 0  (\bar{I}_{2}^{(b)} \approx 0)$ while at the same time $\bar{I}_{2}^{(b)} \neq 0  (\bar{I}_{2}^{(f)} 
\neq 0)$. In other words, for a specific set of parameters $(\Delta T^*,\Omega^*)$ the photonic circuit operates as a perfect 
diode for thermal radiation i.e. $|{\cal R}(\Delta T^*,\Omega^*)|\approx 1$. 

\begin{figure}
\begin{center}
\includegraphics[width=0.5\textwidth]{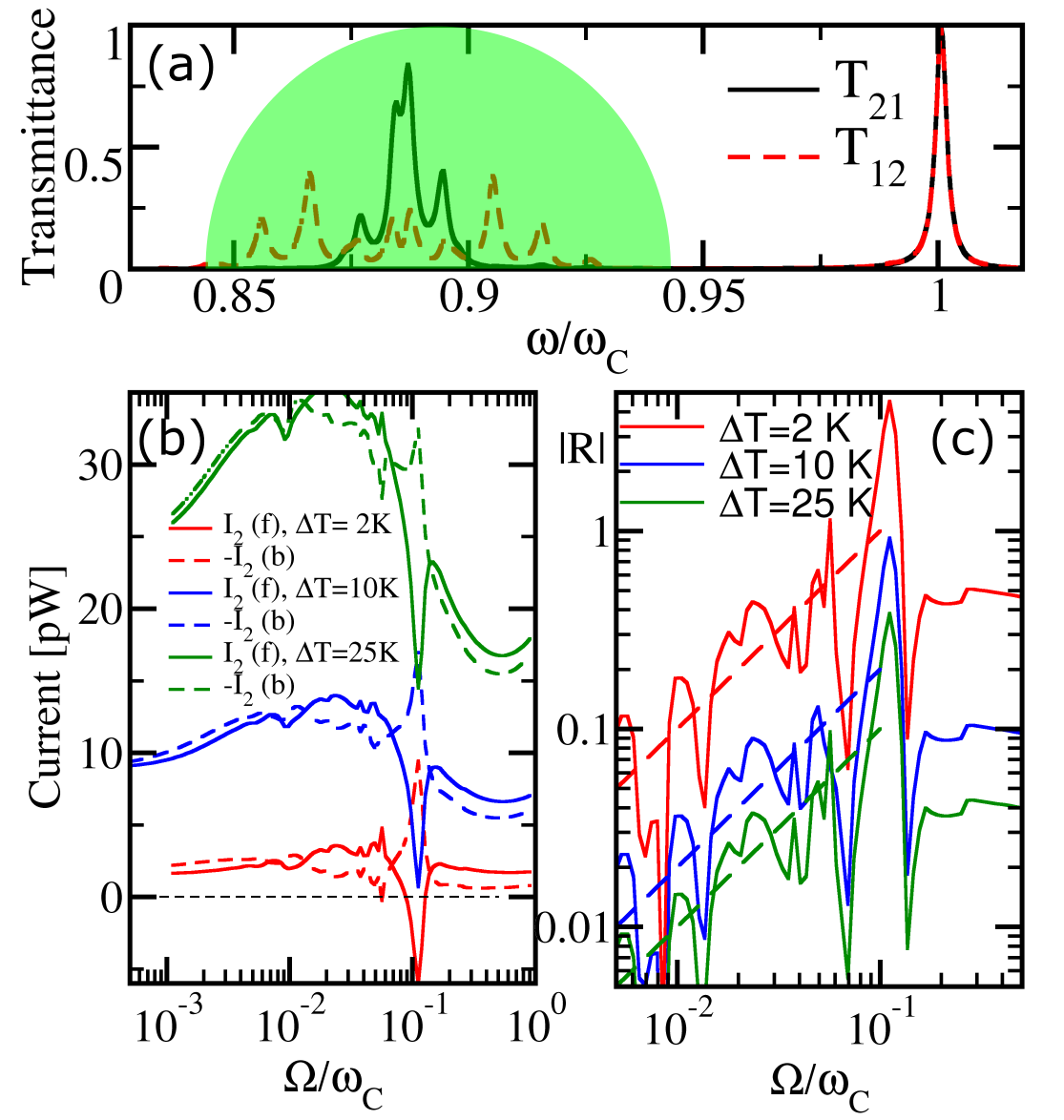}
\caption{(a) Transmittance spectrum ${\cal T}_{\alpha,\beta}^F(\omega)$ for the photonic circuit of Eq. (\ref{H0}) showing 
a nonreciprocal behavior around $\omega\sim 0.88\omega_C$. The driving frequency is $\Omega=0.01\omega_C, k=
0.04\omega_C, \delta_0=0.015\omega_C, \gamma_1=\gamma_2\approx0.004\omega_C$, and $\gamma_3=0$. The green 
area describes the engineered emission spectrum given by Eq. (\ref{filter}). (b) The radiative currents Eq. (\ref{current}) vs. 
$\Omega$ for three representative temperature gradients $\Delta T$. (c) The corresponding (absolute value) rectification 
parameter $|{\cal R}|$ vs. $\Omega$. The dashed lines represents the linear function $|R|=(\alpha/\Delta T)\times \Omega$ 
with $\alpha=20$ given by the best fit.
}
\label{fig2}
\end{center}
\end{figure}

Substituting in Eq. (\ref{eqR}) the results for $\bar{I}_{2}^{(f/b)}$, we find that for $\Omega,\Delta T/T_0
\rightarrow 0$ the rectification parameter is
\begin{equation}
\label{eqR2}
{\cal R}(\Delta T, \Omega)\approx { {\hat {\cal P }} \over {\hat {\cal T}}}{\Omega/(2\pi)\over (\Delta T/T_0)} 
\end{equation}
indicating that thermal rectification increases proportionally to the modulation frequency $\Omega$ and inversely  
proportional to the temperature gradient. The former is responsible for inducing non-reciprocal transport ${\cal 
T}_{\alpha,\beta}^F(\omega)\ne  {\cal T}_{\beta,\alpha}^F(\omega)$ and a pumped current, while the latter controls 
the bias current. From Eq. (\ref{eqR2}) we also conclude that a way to enhance the rectification efficiency is by 
reducing the weighted instantaneous transmittance $\hat{\cal T}$. This goal can be achieved by confining the 
frequency integration in Eq. (\ref{current3}) via a filtering function $\phi(\omega)$ of the emission spectrum of the 
reservoirs. Of course, the filtering process must maintain the frequency range for which the Floquet transmittance 
is non-reciprocal.

%===================================================================================

{\it CMT modeling--} We consider the photonic circuit of Figs. \ref{fig1}a,b described by the effective 
Hamiltonian $H_0$ 
\begin{equation}
H_0=
 \begin{pmatrix}
   \omega_1 & k_{12} & k_{13}\\
  k_{21}  & \omega_2 & k_{23} \\
  k_{31}  & k_{32}  & \omega_3
 \end{pmatrix} 
\label{H0}
\end{equation}
where $k_{nm}=k_{mn}=k$ is the evanescent coupling between the resonators. In the absence of any modulation $\omega_n=\omega_0$, and due to rotational symmetry, the 
system has two degenerate right/left- handed modes $(1,e^{\pm2i\pi/3},e^{\pm4i\pi/3})^T/\sqrt{3}$ with frequency $\omega_{L(R)}=\omega_0-k$ and a mode $(1,1,1)^T/\sqrt{3}$ with frequency $\omega_C=\omega_0+2k$. 

The situation is different in the presence of periodic modulations \cite{SA17,CATSAL18,WMDWSF20}. Guided by previous 
Floquet engineering studies performed in the scattering framework \cite{LSK18,LK19} we have implemented a driving 
protocol that involves the time-modulation of the $n=2,3$-resonators with $\omega_n=\omega_0-\delta_0 \left[ \cos
(\Omega t+\phi_n)+ \cos(\Omega t+\phi_{0}) \right]$, combined with the driving of the coupling constant $k_{23}=k_{32}
=k + \delta_0 \cos(\Omega t+\phi_{0})$. This scheme assumes that the $n=1$-resonator remains undriven i.e. 
$\omega_1=\omega_0$. In this case, the degeneracy of the two counter-rotating modes is lifted and the transmittance 
demonstrates a pronounced non-reciprocal behavior ${\cal T}_{1,2}^F\neq T_{2,1}^F$ around $\omega \approx 
\omega_{L(R)}$ that is maximized by an appropriate choice of the phasors  $\phi_0=0$, $\phi_2=+\pi/2$, and 
$\phi_3=-\pi/2$ (see Fig. \ref{fig2}a) \cite{LK19}. Finally, the modulated coupling introduces an extra non-diagonal 
element in the dissipation matrix $\Gamma$ which becomes 
$\Gamma_{n m} = (\gamma_\alpha \delta_{n,\alpha} 
 - 2 \dot \omega_{n} (t)/\omega_0)\delta_{n,m} - ( 2\dot k(t) /\omega_0) (\delta_{n,2}\delta_{m,3} + \delta_{n,3}\delta_{m,2} )$, 
with $\gamma_3=0$.

In Fig. \ref{fig2}b, we report the currents $\bar{I}_{\alpha=2}^{(f/b)}$ calculated using Eq. (\ref{Floquet_S}) for 
three different temperature gradients $\Delta T$. We observe that as $\Omega$ increases, the radiated current 
becomes non-reciprocal $\bar{I}_2^{(f)} \ne -\bar{I}_2^{(b)}$. The associated rectification parameter ${\cal R}$ 
is shown in Fig. \ref{fig2}c. We find that for small $\Omega$ (and temperature gradients $\Delta T/T_C\ll1$) it 
increases linearly with the modulation frequency and it is inversely proportional to the temperature gradient 
$\Delta T$, in agreement with Eq. (\ref{eqR2}). 

\begin{figure}
\begin{center}
\includegraphics[width=0.5\textwidth]{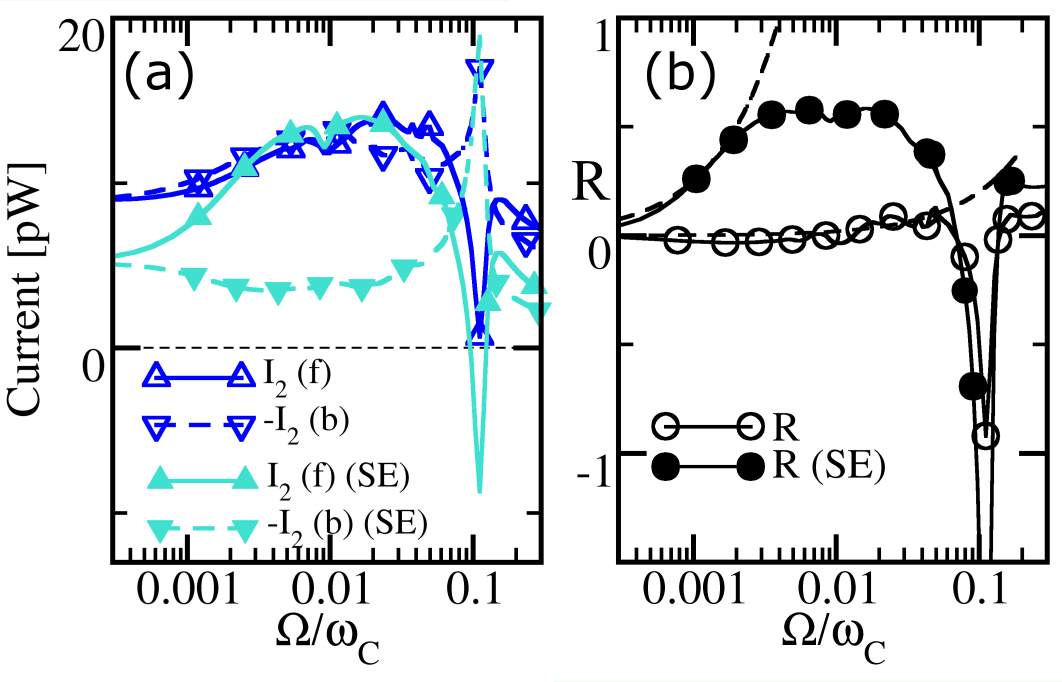}
\caption{(a) The currents Eq. (\ref{current}) vs. $\Omega$ are calculated with (filled symbols) and without 
(open symbols) spectral engineering (SE) for a forward/backward $\bar{I}_2^{(f/b)}$ configuration. (b) The rectification 
${\cal R}$ vs. $\Omega$ for temperature gradient $\Delta T=10K$. Other parameters are as in Fig. 
\ref{fig2}. The black dashed lines indicate a function $R=\alpha \times \Omega$ with best fitting values $\alpha=2 ~(250)$ 
for the unfiltered (filtered) circuit.
}
\label{fig3}
\end{center}
\end{figure}

For the temperature gradient $\Delta T^*=10K$, one can achieve perfect isolation in the forward configuration i.e. 
$I_{2}^{(f)}=0$ while $\bar{I}_2^{(b)}\neq 0$. The associated driving frequency for which the bias current in the 
forward configuration balances the pumped current is $\Omega^*\approx \omega_C-\omega_{L(R)}\approx 3 k$. 
The latter corresponds to a resonant driving that promotes transitions between the frequency domain around 
$\omega\approx \omega_C$, where transport is reciprocal ${\cal T}_{12}^F = {\cal T}_{21}^F$, and the domain 
$\omega\approx\omega_{L/R}$ where ${\cal T}_{12}^F \neq {\cal T}_{21}^F$. For smaller $\Delta T=2K$, the 
biased current in the forward configuration $\bar{I}_{2,b}^{(f)}\sim \Delta T$ is smaller (in magnitude) than the 
pumped current $\bar{I}_{2,p}^{(f)}\sim \Omega$, thus leading to a total emitted radiation from the cold reservoir 
i.e. the circuit operates as a ``refrigerator" with $|{\cal R}|>1$, see Fig. \ref{fig3}. 

Next, we engineered the emission spectrum in a 
way that it excludes the reciprocal frequency range around $\omega\approx \omega_C$ and enforces emission 
in the range where non-reciprocity is maximum. To this end, we have incorporated in Eq. (\ref{bath}) the following 
filtering function 
\begin{equation}
\label{filter}
\phi(\omega)=\Re \left\{ \sqrt{1 - [(\omega-\omega^*)/(b \omega_C)]^2}\right\} 
\end{equation}
where $\omega^*/\omega_C \approx 0.88$ is the frequency around which the transmittance is nonreciprocal and 
$b= 0.05$ is the spectral width of the filtering function. In Figs. \ref{fig3}a,b we report the resulting radiative 
currents and rectification parameter ${\cal R}(\Omega)$ for $\Delta T=10K$. Comparison 
with the unfiltered reservoirs $\phi(\omega)=1$ indicates that the spectrally engineered 
reservoirs lead to a superior rectification. As in the unfiltered case, also here the rectification ${\cal R}\sim \Omega$ 
in the small $\Omega$-regime -- albeit the linear coefficient is much larger (see dashed lines), in agreement with 
the expectations from Eq. (\ref{eqR2}).

%===========================================================================

{\it Electronic Circuit Implementation -} We further validated our results by performing time-domain simulations for 
a realistic electronic circuit (Figs. \ref{fig1}c,d). The latter has been designed using a mapping between the effective 
CMT Hamiltonian Eq. (\ref{H0}) and the circuit's dynamical equations \cite{suppl}. The circuit consists of three LC 
resonators, with identical (and constant) inductances $L$. Modulation in the frequency of the $n=2,3$ LC resonators 
is achieved by changing in time their capacitances as $C_n(t)=C[ 1+  \delta \cos(\Omega t+\varphi_n) ]$. The LC 
elements are capacitively coupled with capacitances $C_c=\kappa C$. The two time-modulated resonators are 
coupled via a modulated capacitance $C_\kappa(t)=C[ \kappa + \delta \cos(\Omega t+\phi_{0}) ]$. Each (undriven) 
resonator supports one resonant mode with frequency $\omega_0=1/\sqrt{LC}=2 \pi 10^9 rad/s$, and resonance 
impedance $z_0=\sqrt{L/C}=70$Ohms. 

The time-dependent voltages at the connection nodes of each resonator $v_{\alpha}(t)$ are driven by synthesized 
noise sources attached to transmission lines (TLs) which are connected to each nodal point. The TLs are introduced 
through their Thevenin equivalent TEM transmission lines with characteristic impedance $Z_0=50 Ohms$. They are 
coupled to the resonators through small capacitances $C_e=\epsilon C$. The noise sources $V_{\alpha}$ are 
synthesized such that 
\begin{equation}
\langle V_{\alpha'}(\omega)V^*_{\alpha}(\omega^{\prime})\rangle=\frac{2Z_0}{\pi}\phi_{\alpha}(\omega)\hbar\omega
\Theta_{\alpha}(\omega)\delta(\omega-\omega^{\prime})\delta_{\alpha,\alpha'} 
\label{control}
\end{equation}    
where $\Theta_{\alpha}(\omega)=k_B T_{\alpha}$ is evaluated at its classical limit and $\phi_{\alpha}(\omega)=
\phi(\omega)$ describes a filtering function.

\begin{figure}
\begin{center}
    \includegraphics[width=0.5\textwidth]{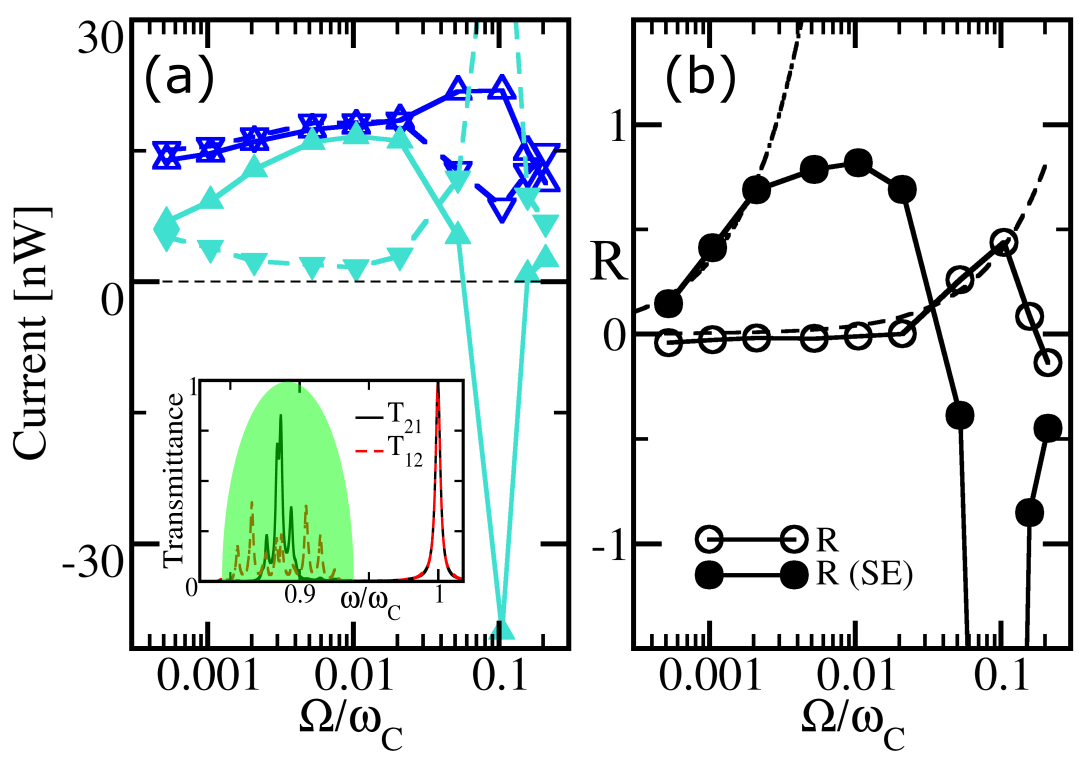}
\end{center}
\caption{Time-domain simulations for the electronic circuit of Figs. \ref{fig1}c,d. (a) The radiative currents Eq. (\ref{Palpha}) 
vs. $\Omega$ are calculated with (filled symbols) and without (open symbols) spectral engineering (SE) for a forward/
backward $\bar{I}_2^{(f/b)}$ configuration. In the inset we show the transmission spectrum. (b) The rectification parameter 
${\cal R}$ vs. $\Omega$ for temperature gradient $\Delta T=10K$. The black dashed lines indicate a function $R=
\alpha \times\Omega$ with $\alpha=4$ and $350$ for the unfiltered and filtered circuit respectively.
We have used the parameters $\kappa=0.1$, $\epsilon=0.1$, $\delta=0.05$, and $T_C=300K$.
}
\label{fig4}
\end{figure}

The net energy current flowing to a transmission line $\alpha$ is evaluated from the time-dependent voltages 
$v_{\alpha}(t)$ and currents $i_{\alpha}(t)$ at the respective nodes,
\begin{equation}
\bar{I}_{\alpha}=\int d\omega {\bar I}_{\alpha}(\omega);\quad
\bar{I}_{\alpha}(\omega)=\frac{\Omega}{2\pi}\int_{t_0}^{t_0+\frac{2\pi}{\Omega}} dt 
\left[ v_{\alpha}(t,\omega) i_{\alpha}(t,\omega) \right], 
\label{Palpha}
\end{equation}
where an average over one modulation cycle is assumed. Moreover, an initial transient $t_0$ has been discarded 
to ensure steady state conditions. The transmittances ${\cal T}_{1,2}, {\cal T}_{2,1}$ are obtained from 
$\bar{I}_{\alpha}(\omega)$ in Eq. (\ref{Palpha}), by setting $V_{\beta^\prime}= 0$, with $\beta^\prime\ne \beta$, 
and normalizing the incident currents to unit power flux, see inset of Fig. \ref{fig4}a. A comparison with the corresponding 
CMT results (Fig. \ref{fig2}a) confirms the efficiency of our modeling. 

In Fig. \ref{fig4}a we compare the thermal radiation for the forward (Fig \ref{fig1}c) and backward (Fig. \ref{fig1}d) 
configurations, in the absence and presence of spectral filtering. For the latter case, we have used the filtering function 
$\phi(\omega)$ of Eq. (\ref{filter}). The currents $\bar{I}_2^{(f/b)}$ are in quantitative agreement with the CMT results. 
Similarly, the rectification parameter ${\cal R}$ for both the unfiltered (open circles) and filtered (filled circles) electronic 
circuits (Fig. \ref{fig4}c) are in agreement with the theoretical predictions of Eq. (\ref{eqR2}). 
We find a linear behavior with $\Omega$ (black dashed lines) with the linear coefficient in the case of spectrally 
engineered baths being two orders larger than the corresponding coefficient found for the unfiltered case. 
 
%================================================================================= 
{\it Conclusion.-} We have unveiled the interplay between pumped currents, associated with Floquet driving, and biased 
currents, associated with the temperature gradient between two reservoirs. When these elements are interlaced with 
judiciously engineered spectral filters of the reservoirs, they lead to extreme non-reciprocal thermal radiation. Our results 
can be used for the design of thermal circulators, and for the identification of efficient refrigeration protocols.

{\it Acknowledgements.-} (LJFA, TK) acknowledge partial support by an ONR Grant No. N00014-16-1-2803, by an 
AFOSR Grant No. FA 9550-14-1-0037 and by an NSF Grants No. EFMA-1641109. The postdoctoral work of (HL) 
at Wesleyan University was supported via grant AFOSR Grant No. FA 9550-14-1-0037.

\newpage

%%%%%%%%%% Merge with supplemental materials %%%%%%%%%%
%%%%%%%%%% Prefix a "S" to all equations, figures, tables and reset the counter %%%%%%%%%%
\setcounter{equation}{0}
\setcounter{figure}{0}
\setcounter{table}{0}
\setcounter{page}{1}
\makeatletter
\renewcommand{\theequation}{S\arabic{equation}}
\renewcommand{\thefigure}{S\arabic{figure}}
\renewcommand{\bibnumfmt}[1]{[S#1]}
\renewcommand{\citenumfont}[1]{S#1}
%%%%%%%%%% Prefix a "S" to all equations, figures, tables and reset the counter %%%%%%%%%% 

\section{Supplemental Material} 
 
\subsection{Energy Current and Floquet Scattering Matrix}

In this section, our goal is to provide a derivation for
the net average energy current 
$\bar{I}_\alpha$
directed toward a heat bath $\alpha$, Eq. (4) of the main text, when a scatterer connecting thermal reservoirs is periodically driven. 
We start by considering the waves $\left| \psi (t) \right \rangle$ inside the scatterer, evolving according to the equation
\begin{equation}
     \frac{d \left| \psi (t) \right \rangle}{dt} = \left[ i  H(t) - \Gamma \right]\left| \psi(t) \right \rangle
     + D^{+} \left| S_{+}(t) \right \rangle.
     \label{eqS1}
\end{equation}
The field amplitude $\left| \psi (t) \right \rangle$ is a result of the excitations $ \left| S_{+}(t) \right \rangle$ coming from 
the heat baths connected to the system through the (frequency-independent) coupling matrix $D$. The time dependent 
Hamiltonian $H(t)$ and the losses $\Gamma=D D^{+} / 2$ not only determine the dynamics of the wave function, but 
also shape the outgoing scattered waves
\begin{equation}
      \left| S_{-}(t) \right \rangle =  - \left| S_{+}(t) \right \rangle 
      + D\left| \psi(t) \right \rangle.
      \label{eqS2}
\end{equation}

These complex quantities in coupled mode theory are represented in the frequency domain through their 
positive frequency component $\left| f(t) \right \rangle=\int_0^\infty d\omega \left| f(\omega) \right \rangle 
e^{i  \omega t}$ and $\left| f(\omega) \right \rangle=\left| f(-\omega) \right \rangle ^* =
\int^{+\infty}_{-\infty} \left| f(t) \right \rangle  e^{-i  \omega t}d t/(2\pi); \omega>0$
where $f$ is $S_{\pm}$ or $\psi$.
The effective Hamiltonian, being periodic in time, results
\begin{equation}
H_{\rm eff}(t)=H_{\rm eff}(t+\frac{2\pi}{\Omega})=H(t) + i  \Gamma=\sum_{m=-\infty}^{+\infty} e^{i  m 
\Omega t} H_Q^{0,(m)},
\end{equation}
where $\Omega$ is the modulation frequency and $H_Q^{0,(m)}=\frac{\Omega}{2\pi}\int_0^{2\pi/\Omega} 
H_{\rm eff}(t) e^{-i  m \Omega t} dt$ is a $N_S\times N_S$ matrix, being $N_S$ the number of modes.
In what follows, we assume that Eq. (\ref{eqS1}) is valid around a resonant frequency $\omega_0$, and $\left| 
\psi (\omega) \right \rangle \rightarrow 0$ when $\omega\gg\omega_0$ or $\omega\ll \omega_0$ and that 
$\Omega \ll \omega_0$. In addition, the thermal excitations coming from bath $\beta$, $S^{+}_{\beta}(\omega)
=\left\langle \beta | S_+ (\omega) \right \rangle $, satisfy the correlation relation
\begin{equation}
\left\langle  [S^{+}_{\beta^\prime}(\omega^\prime)]^* S^{+}_{\beta}(\omega) \right \rangle = 
\frac{\tilde{\Theta}_\beta(\omega)}{2\pi}\delta _{\beta^\prime,\beta} \delta(\omega-\omega^\prime),     
\label{eqS4}
\end{equation}
where $\tilde{\Theta}_\beta(\omega)= \hbar \omega[\exp(\hbar\omega/(k_B T_\beta))-1]^{-1}$
being $k_B$ and $T_\beta$ the Boltzmann constant and temperature of reservoir $\beta$, respectively.
Therefore, from Eq. (\ref{eqS1}) we have
\begin{equation}
\omega \left| \psi (\omega) \right \rangle = \sum_m H_Q^{0,(m)} \left| \psi (\omega-m\Omega) \right \rangle
-i  D^T \left| S_+ (\omega) \right \rangle.
\label{eqS5}
\end{equation}
We can turn Eq. (\ref{eqS5}) into a matrix equation in an extended space
\begin{equation}
\left([\hat{\omega}]-\hat{H}_Q \right)  \vec{\psi} (\hat{\omega})  = -i  [D]^T  \vec{S}_+ (\hat{\omega}), 
\label{eqS6}
\end{equation}
with the definition of the following quantities. The block matrix $\hat{H}_Q=H_Q^0-[n\Omega]$, 
where $H_Q^0$ and $[n\Omega]$ are block matrices, whose blocks are $(H_Q^0)_{p,q}=H_Q^{0, (q-p)}$ and 
$[n\Omega]={\rm diag}\left\{ \cdots, \Omega I_{N_S}, 0, -\Omega I_{N_S}  , \cdots \right\}$, respectively. Here,
$I_{N_S}$ is the $N_S \times N_S $ identity matrix and the notation $[A]$ represents 
a block diagonal matrix whose blocks are $A I_{N_S}$. 
We denote the frequency as $\hat{\omega}$ when its range is restricted to $\hat{\omega}\in [\omega_0-\Omega/2 , \omega_0+\Omega/2 )$.
Finally, we define the vectors 
$\vec{f} = [ \cdots, \left| f (\omega_{+1}) \right \rangle , \left| f (\omega_0) \right \rangle , \left| f (\omega_{-1}) 
\right \rangle, \cdots ]^T$, with $f$ being, as before, $S_{\pm}$ or $\psi$, and where $\omega_n=\hat{\omega}+
n\Omega$.

Interestingly, Eq. (\ref{eqS6}) allow us to express the wave function vector in a physically meaningful form, 
\begin{equation}
\vec{\psi} (\hat{\omega})  = -i   G^F  [D]^T  \vec{S}_+ (\hat{\omega}), 
\label{eqS7}
\end{equation}
which evidences that the excitations introduced by the thermal baths, $\vec{S}_+ (\hat{\omega})$,
are propagated through the system in the extended dimension, and hence scattered to other frequencies. 
The propagator,
\begin{equation}
G^F=\left([\hat{\omega}]-\hat{H}_Q \right) ^{-1},
\end{equation}
is nothing else than the Green's function of the extended space, or also called Floquet Green's function, which allow us to
find the outgoing scattered fields for given incident waves $\vec{S}_+ (\hat{\omega})$.

The above mentioned outgoing scattered field
% , $\left| S_{-}(\omega) \right \rangle =  - \left| S_{+}(\omega) \right \rangle + D\left| \psi(\omega) \right \rangle$, now 
can be readily found by using the Fourier transform of Eq. (\ref{eqS2}) in the extended space and Eq. (\ref{eqS7}),
\begin{eqnarray}
\vec{S}_- (\hat{\omega}) &=& -\vec{S}_+ (\hat{\omega})  + [D] \vec{\psi} (\hat{\omega}), \notag \\
 &=& \left( -I -i  [D] G^F  [D]^T \right) \vec{S}_+ (\hat{\omega}),
\end{eqnarray}
where $I=[1]$. Here, we identify the term inside the parenthesis as the Floquet Scattering matrix
\begin{equation}
{\cal S}^F=-I -i  [D] G^F  [D]^T,
\end{equation}
and this allow us to find the scattered field going out of the system toward lead $\alpha$
\begin{equation}
S_{\alpha n}^-(\hat{\omega}) = \sum_{\beta,m} {\cal S}^F_{\alpha n,\beta m} (\hat{\omega}) S^+_{\beta m}(\hat{\omega}),
\label{eqSmin1}
\end{equation}
where $S^\pm_{\beta}(\omega_m)=S^\pm_{\beta m}(\hat{\omega})=\langle \beta | S_\pm (\hat{\omega} + m\Omega)\rangle$.
Notice that the element ${\cal S}^F_{\alpha n,\beta m} (\hat{\omega})$ indicates that radiation coming from lead $\beta$ at frequency 
$\omega_m=\hat{\omega}+m \Omega$ leaves the system toward lead $\alpha$ with 
frequency $\omega_n=\hat{\omega}+n \Omega$. Then, in order to highlight the incident and outgoing frequencies, we will also use the notation
${\cal S}^F_{\alpha,\beta} (\omega_n , \omega_m)\equiv{\cal S}^F_{\alpha n,\beta m} (\hat{\omega})$.

The net average energy current
going out of the system toward the reservoir $\alpha$ is
\begin{equation}
\bar{I}^{out}_\alpha = \frac{\Omega}{2\pi} \int _0^{2\pi/\Omega} dt
\left\langle | S^-_\alpha (t) |^2 \right\rangle,
\label{Iout1}
\end{equation}
where the outgoing field in the frequency domain can be written as 
\begin{equation}
S^-_\alpha (t)=\sum_n \int _{\omega_0 - \frac{\Omega}{2}}^{\omega_0 + \frac{\Omega}{2}} S_{\alpha n}^- (\hat{\omega}) e^{i  (\hat{\omega} + n\Omega)t } d\hat{\omega}.
\label{eqSmin2}
\end{equation}
Introducing Eq. (\ref{eqSmin2}) into Eq. (\ref{Iout1}) leads us to evaluate the correlation for the outgoing scattered fields, which read
\begin{eqnarray}
\left\langle \left(  S^-_{\alpha n}(\hat{\omega})  \right)^*  
S^-_{\alpha n^\prime}(\hat{\omega^\prime}) \right\rangle 
&=&\sum_{\beta, m} \left({\cal S}^F_{\alpha n,\beta m}(\hat{\omega})\right)^*
{\cal S}^F_{\alpha n^\prime,\beta m}(\hat{\omega}) \times \notag \\
 &\times&
\frac{\tilde{\Theta}_\beta(\hat{\omega} + m\Omega)}{2\pi}   \delta(\hat{\omega}-\hat{\omega}^\prime).
\end{eqnarray}
Here we have used Eqs. (\ref{eqSmin1}), (\ref{eqSmin2}), and 
the correlation relations for the incident radiation 
\begin{equation}
\left\langle \left(  S^+_{\beta m}(\hat{\omega})  \right)^*  
S^+_{\beta^\prime m^\prime}(\hat{\omega^\prime}) \right\rangle 
=\frac{\tilde{\Theta}_\beta(\hat{\omega} + m\Omega)}{2\pi}  \delta_{\beta,\beta^\prime} \delta_{m,m^\prime} \delta(\hat{\omega}-\hat{\omega}^\prime),
\label{corrSplus}
\end{equation}
which follow from the properties of the thermal reservoirs, Eq. (\ref{eqS4}).
Therefore, we obtain 
\begin{eqnarray}
\bar{I}^{out}_\alpha &=& \sum_n \int _{\omega_0 - \frac{\Omega}{2}}^{\omega_0 + \frac{\Omega}{2}} \frac{d\hat{\omega}}{2\pi} \sum_{\beta,m} 
\left| {\cal S}^F_{\alpha n,\beta m } (\hat{\omega}) \right|^2 \tilde{\Theta}_\beta (\hat{\omega} + m\Omega)  \notag \\
&=&  \int _{0}^{\infty} \frac{d\omega}{2\pi} \sum_{\beta,m} 
\left| {\cal S}^F_{\alpha,\beta } (\omega,\omega_m) \right|^2 \tilde{\Theta}_\beta (\omega_m)
\label{Iout2}
\end{eqnarray}
where we have used $\frac{\Omega}{2\pi} \int_0^{2\pi/\Omega} dt e^{i  (n^\prime-n)\Omega t}=\delta_{n^\prime,n}$.
Notice that in Eq. (\ref{Iout2}) we have restored the integration over the whole frequency range by using the incident-outgoing frequency notation for ${\cal S}^F$.

Finally, we can evaluate the net average energy current
going toward reservoir $\alpha$
\begin{equation}
 \bar{I}_\alpha = - \bar{I}^{in}_\alpha + \bar{I}^{out}_\alpha ,
 \label{net_current}
\end{equation}
where the incident energy current
$
\bar{I}^{in}_\alpha = \frac{\Omega}{2\pi} \int _0^{2\pi/\Omega} dt \left\langle | S^+_\alpha (t) |^2 \right\rangle = \int _0^{\infty} \frac{d\omega}{2\pi} \tilde{\Theta}_\alpha (\omega).
$
Finally, using Eq. (\ref{Iout2}) and shifting $\omega_m \rightarrow \omega$, we obtain
\begin{equation}
\bar{I}_\alpha = \int _0^{\infty} \frac{d\omega}{2\pi}
\sum_{\beta n} \left[ -\delta_{\beta, \alpha} \delta_{n,0} + 
\left| {\cal S}^F_{\alpha \beta} (\omega_n,\omega)  \right|^2 \right]\tilde{\Theta}_\beta (\omega),
\end{equation}
which demonstrates Eq. (4) of the main text.
Notice that, $\bar{I}_\alpha$ is defined as positive when the  current is going toward the reservoir $\alpha$.

\subsection{Energy Current in the adiabatic limit.}
Here, we provide expressions for the average energy current in the adiabatic limit, $\Omega\rightarrow 0$, without involving the classical limit and small temperature gradient approximation. Further details will be given in a future publication\cite{NFLK20}. The average current at a lead $2$ can be separate in two contributions, as in Eq. (6) of the main text,
\begin{equation}
   \bar{I}_{2} \approx \bar{I}_{2,b} + \bar{I}_{2,p}.
\end{equation}
Here, the bias current reads
\begin{equation}
\bar{I}_{2,b}=  \int \frac{d\omega}{2\pi} \hbar \omega {\bar {\cal T }}(\omega) 
\left( \Theta_1(\omega)-\Theta_2(\omega) \right),
\end{equation}
where the transmittance $\bar{{\cal T}}(\omega)=\frac{\Omega}{2\pi}\int dt |S^t_{21}(\omega)|^2$ is averaged over one cycle, being $S^t$ the instantaneous scattering matrix, and we consider instantaneous reciprocal transport, i.e. $S^t_{21}(\omega)=S^t_{12}(\omega)$.
The current associated with the modulation of the scatterer is the pumped current, which evaluated at lead 2 reads
\begin{eqnarray}
\bar{I}_{2,p}^{(f)} &\approx& \frac{\Omega}{2\pi} \int \frac{d\omega}{2\pi} 
\left\{ \hbar \omega \Theta_{2}(\omega) \frac{\partial P_{22}}{\partial \omega}+ \right. \label{S_pump} \\
&+& P_{22}(\omega) \frac{\Theta_{1}(\omega)+\Theta_{2}(\omega)}{2} + \nonumber \\
&+& \left.  \left[A_{22} - A_{21}\right]  \frac{\Theta_{1}(\omega)-\Theta_{2}(\omega)}{2} \right\} \nonumber
\end{eqnarray}
where $A_{\alpha,\beta}(\omega)= \im \int dt \frac{d S^t_{\alpha,\beta}}{dt}(S^t_{\alpha,\beta})^{*}$, and 
$P_{22}(\omega)= \im \int dt \left( \frac{d S^t}{dt}(S^t)^{\dagger} \right)_{2,2} = A_{22}+A_{21}$.
In the classical limit, where $\Theta_{\alpha}(\omega)\approx k_B T_\alpha /(\hbar \omega) $,
integration of the first term of Eq. (\ref{S_pump}) results proportional to 
$\left[P_{22}(\omega)\right]^\infty_0\approx0$, while the third term is of order ${\cal O} (\Omega \Delta T/T_0)$.

\subsection{Decimation Procedures, Effective Hamiltonians, and Green's Functions}

The main difficulty in the computation of ${\cal S}^F$ in Eq. (3) of the main text is associated with the evaluation 
of the Floquet Green's function $G^F=([\omega] -\hat{H}_Q)^{-1}$ which requires the inversion of the matrix 
$([\omega] -\hat{H}_Q)$, whose rank is in principle infinite involving all Floquet channels $n\in(-\infty,+\infty)$. 
Approximate results can be obtained through truncations of the Floquet space to $n\in[-N_F,N_F]$, where 
reliable results require $N_F$ to be large, slowing down the calculation.

Of course, there are cases, like for adiabatic \cite{LAESK19} 
and for high modulation frequencies \cite{LSK18,LK19} or for a simple two level Rabi driving schemes \cite{BLF20}, 
where $G^F$ and subsequently $\bar{I}_{\alpha}$ are easily calculated. In most general scenarios, however, one 
needs to consider many Floquet channels in order to obtain an accurate description of the scattering process. We 
have tackled this difficulty by employing a decimation technique borrowed from the field of molecular 
electronics \cite{DABD83,PM01,CFBNP14}. By utilizing the block diagonal structure of the Hamiltonian in the Floquet
-Hilbert space, a matrix continued fraction expansion \cite{CFBNP14} allows the calculation of $G^F$ via an iteration 
relation connecting blocks $n$ and $m$. 

Floquet Hamiltonians $\hat{H}_Q$ have typically a block structure. In particular, for 
simple driving schemes (few harmonics in the Fourier expansion of the effective Hamiltonian)
$\hat{H}_Q$ is block tridiagonal and thus several nondiagonal blocks are zeros.
Here, we take advantage of this structure and we perform efficient calculation of $G^F$
by means of the {\it decimation} procedures, inspired in the renormalization group techniques
of statistical mechanics \cite{DABD83}, and widely utilized in Condensed Matter \cite{CFBNP14}
and Molecular Electronics \cite{PM01}. Here we will show the basics of this technique, 
and we parallel the approach given in Refs. \cite{PM01,CFBNP14}.

%------------------------------------------------------

The decimation procedures recursively reduce a general $N\times N$ Hamiltonian
into another of lower rank by decreasing the number of degrees of freedom, 
without altering its physical properties. As a result, the method utilizes ${\cal O}(N)$ operations 
instead of ${\cal O}(N^2)$ required by the matrix inversion \cite{CFBNP14} and allow us to deal 
with complex driving schemes which are intractable by any other method.
For instructive purposes, let us consider a block tridiagonal Hamiltonian ${\mathbb H}$ such that 
it satisfies the equation
\begin{equation}
\left[ 
\begin{array}{ccc}
\omega -{\mathbb E}_{1} & -{\mathbb V}_{12} & {\mathbb O} \\ 
-{\mathbb V}_{21} & \omega -{\mathbb E}_{2} & -{\mathbb V}_{23} \\ 
{\mathbb O} & -{\mathbb V}_{32} & \omega -{\mathbb E}_{3}
\end{array}%
\right] \left( 
\begin{array}{c}
{\vec u}_{1} \\ 
{\vec u}_{2} \\ 
{\vec u}_{3}%
\end{array}%
\right) =\left[ \omega -\mathbb{H}\right] {\vec u}={\vec 0},
\label{eq:Ham}
\end{equation}%
where the corresponding identity matrices multiplying $\omega$ are implicit.
From the middle (block) equation, we can isolate $\vec{u}_2$ and \textit{decimate} it, 
leading to the equations
\begin{equation}
\left[ 
\begin{array}{cc}
\omega -\overline{\mathbb E}_{1} & -\overline{\mathbb V}_{13} \\ 
-\overline{\mathbb V}_{31} & \omega -\overline{\mathbb E}_{3}%
\end{array}%
\right] \left( 
\begin{array}{c}
\vec{u}_{1} \\ 
\vec{u}_{3}%
\end{array}%
\right)  \label{H_S2eff} \\
=[\omega -\mathbb{H}_{\mathrm{eff.}}]\vec{u}=0.
\end{equation}%
Here, the blocks have been renormalized hiding the nonlinear dependence on $\omega$:
\begin{eqnarray}
\overline{\mathbb E}_{1}&=&{\mathbb E}_{1}+\Sigma _{1}(\omega )={\mathbb E}_{1}
+{\mathbb V}_{12}\left( \omega -{\mathbb E}_2\right)^{-1} {\mathbb V}_{21}, \notag \\ 
\overline{\mathbb E}_{3}&=&{\mathbb E}_{3}+\Sigma _{3}(\omega )={\mathbb E}_{3}
+{\mathbb V}_{32}\left( \omega -{\mathbb E}_2\right)^{-1} {\mathbb V}_{23}, \notag \\ 
\overline{\mathbb V}_{13} &=&{\mathbb V}_{12}\dfrac{1}{\omega -{\mathbb E}_{2}} {\mathbb V}_{23}; 
\quad \overline{\mathbb V}_{31} ={\mathbb V}_{32}\dfrac{1}{\omega -{\mathbb E}_{2}} {\mathbb V}_{21}.%
\label{eq:decim-ex}
\end{eqnarray}%
The terms $\Sigma _{j}(\omega )$, known as self-energies,
account for the frequency (energy) shifts due to the coupling with the
decimated state. 
Notice that now, there are effective coupling elements $\overline{\mathbb V}_{13(31)}$
between blocks $1$ and $3$ accounting for the interaction of those blocks mediated 
by the decimated block $2$.
Importantly, the nonlinear dependence on $\omega $
codifies all information on the steady state scattering as well as on the dynamics.
For instance, equation \ref{H_S2eff}
gives the exact spectrum of the whole system.

Now, let us come back to eq. \ref{eq:Ham} and decimate block 3 and then 2. 
According to Eq. (\ref{eq:decim-ex}), now we have only  block 1, which is renormalized as 
\begin{eqnarray}
 \tilde{\mathbb E}_{1}&=& {\mathbb E}_{1}+\Sigma _{1}^{(3)}(\omega ) \notag \\
  &=&{\mathbb E}_{1} + {\mathbb V}_{12}  \left( \omega -{\mathbb E}_2 - \Sigma_{2}^{(3)}  \right)^{-1}  {\mathbb V}_{21}  \label{E1}\\
  &=&  {\mathbb E}_{1} + {\mathbb V}_{12}  \left[ \omega -{\mathbb E}_2 - {\mathbb V}_{23}   ( \omega -{\mathbb E}_3 )^{-1} 
  {\mathbb V}_{32} \right]^{-1}  {\mathbb V}_{21} \notag
\end{eqnarray}
Here, we have introduced the notation $\Sigma_{n}^{(m)} $
to indicate the correction to block $n$ due to the decimation of all blocks 
between $n$ and $m$, with $m$ included. 
We highlight that the order of the decimation protocol does not affect $\Sigma_{n}^{(m)}$, 
which is obtained as ``matrix continued fractions'' \cite{CFBNP14}.

%------------------------------------------------------

The recursive structure of the self-energy $\Sigma_{n}^{(m)}$, e.g. as shown in Eq. (\ref{E1}), 
can be used to efficiently reduce Hamiltonians of arbitrary dimensions.
In particular, the Floquet Hamiltonian can be decimated into two blocks,
with labels $n$ and $m$, resulting in
\begin{eqnarray}
\tilde{\mathbb{E}}_{n}&=&\mathbb{E}_{n}+\Sigma _{n}^{(1)}+\Sigma _{n}^{(m)} \notag\\ 
\tilde{\mathbb{E}}_{m}&=&\mathbb{E}_{m}+\Sigma _{m}^{(N)}+\Sigma _{m}^{(n)} \notag\\ 
\tilde{\mathbb{V}}_{n,m}&=&\tilde{\mathbb{V}}_{n,m-1}(\omega-\mathbb{E}_{m}-\Sigma _{m}^{(n)})^{-1}\mathbb{V}_{m-1,m}
\end{eqnarray}%
where 
\begin{eqnarray}
\Sigma _{n}^{(m)}&=&\left[ \mathbb{V}_{n,n+1}\left( \omega 
-\mathbb{E}_{n+1}-\Sigma _{n+1}^{(m)}\right) ^{-1}\right] 
\mathbb{V}_{n+1,n} \notag \\ 
\Sigma _{m}^{(n)}&=&\left[ \mathbb{V}_{m,m-1}\left( \omega 
-\mathbb{E}_{m-1}-\Sigma _{m-1}^{(n)}\right) ^{-1}\right] 
\mathbb{V}_{m-1,m}
\end{eqnarray}%
for $m>n$. We have assumed block tridiagonal matrices, but the procedure can be straightforwardly generalized.

Now, utilizing this procedure, we can address our initial question 
by obtaining the block element of the total Green's function connecting blocks $n$ and $m$ from
\begin{equation}
\left[ 
\begin{array}{cc}
\mathbb{G}_{nn} & \mathbb{G}_{nm} \\ 
\mathbb{G}_{mn} & \mathbb{G}_{mm}%
\end{array}%
\right] =\left[ 
\begin{array}{cc}
\omega -\tilde{\mathbb{E}}_{n} & -\tilde{\mathbb{V}}_{nm} \\ 
-\tilde{\mathbb{V}}_{mn} & \omega -\tilde{\mathbb{E}}_{m}%
\end{array}%
\right] ^{-1}.
\end{equation}%
The inversion of the matrix can be performed resorting to the block-inversion matrix
% \begin{widetext}
\begin{eqnarray}
& &
\left[ 
\begin{array}{cc}
\mathbb{A} & \mathbb{B} \\ 
\mathbb{C} & \mathbb{D}%
\end{array}%
\right] ^{-1}= \label{eq:BlockInvert}\\
&=& \left[ 
\begin{array}{cc}
(\mathbb{A}-\mathbb{BD}^{-1}\mathbb{C})^{-1} & 
-(\mathbb{A}-\mathbb{BD}^{-1}\mathbb{C})^{-1}\mathbb{B}\mathbb{D}^{-1} \\ 
-(\mathbb{D}-\mathbb{CA}^{-1}\mathbb{B})^{-1}\mathbb{C}\mathbb{A}^{-1} & 
(\mathbb{D}-\mathbb{CA}^{-1}\mathbb{B})^{-1}%
\end{array}%
\right] ,  \notag
\end{eqnarray}%
% \end{widetext}
which requires the existence of the inverses of matrices 
$\mathbb{D}$, $\mathbb{A}$, $(\mathbb{D}-\mathbb{CA}^{-1}\mathbb{B})$, and
$(\mathbb{A}-\mathbb{BD}^{-1}\mathbb{C})$. In our case, all of them exist.
Therefore we have, 
\begin{equation}
\begin{array}{c}
\mathbb{G}_{nn}=\left[ (\omega -{\mathbb{E}}_{n})-\Sigma _{n}^{(1)}
-\Sigma _{n}^{(N)}\right] ^{-1}, \\ 
\mathbb{G}_{mm}=\left[ (\omega -{\mathbb{E}}_{m})-
\Sigma _{m}^{(1)}-\Sigma _{m}^{(N)}\right] ^{-1}, \\ 
\mathbb{G}_{nm}=\mathbb{G}_{nn}\left[ \tilde{\mathbb{V}}_{nm}(\omega 
-\tilde{\mathbb{E}}_{m})^{-1}\right] , \\ 
\mathbb{G}_{mn}=\mathbb{G}_{mm}\left[ \tilde{\mathbb{V}}_{mn}(\omega 
-\tilde{\mathbb{E}}_{n})^{-1}\right] .%
\end{array}
\label{eq:Green-blocksF}
\end{equation}%
These equations allow the calculation of the Green's functions avoiding the inversion 
of the full Floquet Hamiltonian matrix.
In Eq. (\ref{eq:Green-blocksF}) the recursive nature of the decimation procedure requires 
$\mathcal{O}(N)$ self energies $\Sigma _{n}^{(1)}$ and $\Sigma_{n}^{(N)}$ for the diagonal elements of $G^F$.
For the non diagonal elements, it is possible to use the self energies already calculated 
for the diagonal ones, highly improving the performance of the method.\cite{CFBNP14}

%================================================================================
%================================================================================

\subsection{Coupled mode theory description of the electrical circuits}

We construct a CMT description for the electrical circuit as shown
in Fig. 1(c,d). As a first step, we analyze a LC resonator using a complex
mode amplitude $\psi$. Specifically, we define the mode amplitude
$\psi\left(t\right)$ to be $\psi\left(t\right)\equiv\sqrt{C/2}\left[v\left(t\right)+\dot{v}\left(t\right)/(j\omega_{0})\right]$
in terms of the node voltage $v\left(t\right)$ and its time derivative
$\dot{v}\left(t\right)$, where $\omega_{0}=1/\sqrt{LC}$ is the resonant
(angular) frequency of the LC resonator. The definition of the mode
amplitude $\psi$ allows us to rewrite the circuit equation, i.e.,
$\frac{d^{2}v\left(t\right)}{dt^{2}}+\omega_{0}^{2}v\left(t\right)=0$,
equivalently as the first-order differential equation $\frac{d}{dt}\psi\left(t\right)=j\omega_{0}\psi\left(t\right)$
or its complex conjugate. At the same time, the mode amplitude $\psi$
is normalized such that $\left|\psi\right|^{2}$ represents the energy
stored in the resonator. Notice that the full degree of freedom i.e.,
$v\left(t\right)$ and $\dot{v}\left(t\right)$, required to specify
the circuits completely at each time, is maintained in the complex-mode
description, since they can be expressed, using the definition of
the amplitude $\psi\left(t\right)$ and its complex conjugate $\psi^{*}\left(t\right)$,
as $v\left(t\right)=\frac{1}{\sqrt{2C}}\left[\psi\left(t\right)+\psi^{*}\left(t\right)\right]$
and $\dot{v}\left(t\right)=j\omega_{0}\frac{1}{\sqrt{2C}}\left[\psi\left(t\right)-\psi^{*}\left(t\right)\right]$.
Nevertheless, when describing the dynamics of circuits, the amplitude
$\psi\left(t\right)$ and its complex conjugate $\psi^{*}\left(t\right)$
are generally not decoupled with each other as seen below.

We proceed to describe the coupling between two (identical) LC resonators
under the complex-mode description. As considered in Fig. 1(c,d), the
coupling between the resonators can be enabled by the capacitor $C_{c}=\kappa C$.
According to Kirchhoff's laws, the circuit equations describing the
coupled LC resonators simply read
\begin{align}
\left(1+\kappa\right)\frac{d^{2}v_{1}}{dt^{2}}-\kappa\frac{d^{2}v_{2}}{dt^{2}}+\omega_{0}^{2}v_{1} & =0,\label{eq: circuit1a}\\
\left(1+\kappa\right)\frac{d^{2}v_{2}}{dt^{2}}-\kappa\frac{d^{2}v_{1}}{dt^{2}}+\omega_{0}^{2}v_{2} & =0\label{eq: circuit1b}
\end{align}
where $v_{n},n=1,2$ are the node voltages of each resonator. Using
the complex-mode representation $\psi_{n}$ for each resonator $n$,
we can rewrite the circuit equations Eq. (\ref{eq: circuit1a}) and
(\ref{eq: circuit1b}) as 
\begin{align}
\frac{d\psi_{1}}{dt} & \approx j\omega_{0}\psi_{1}-\frac{j\omega_{0}}{2}\kappa\left(\psi_{1}-\psi_{2}+\psi_{1}^{*}-\psi_{2}^{*}\right)\label{eq: C1}\\
\frac{d\psi_{2}}{dt} & \approx j\omega_{0}\psi_{2}-\frac{j\omega_{0}}{2}\kappa\left(\psi_{2}-\psi_{1}+\psi_{2}^{*}-\psi_{1}^{*}\right)\label{eq: C2}
\end{align}
when assuming $\kappa\rightarrow0$. Furthermore, under the rotating-wave
approximation enabled by the weak coupling $\kappa\rightarrow0$,
we can simplify the Eqs. (\ref{eq: C1}) and (\ref{eq: C2}) further
by decoupling $\psi_{n}$ with their complex conjugates $\psi_{n}^{*}$
to get a coupled mode form 
\begin{align}
\frac{d\psi_{1}}{dt} & \approx j\omega_{0}\left(1-\frac{1}{2}\kappa\right)\psi_{1}+\frac{j\omega_{0}}{2}\kappa\psi_{2}\\
\frac{d\psi_{2}}{dt} & \approx\frac{j\omega_{0}}{2}\kappa\psi_{1}+j\omega_{0}\left(1-\frac{1}{2}\kappa\right)\psi_{2}.
\end{align}
Clearly, the capacitive coupling $C_{c}=\kappa C$ shifts the resonant
frequency of each resonators in addition to coupling the two resonators.

The effects of the external capacitive coupling $C_{e}=\epsilon C$
between a transmission line (TL) and a LC resonator can be examined
similarly. Before that, we need to define the complex wave amplitude
$S^{\pm}$ for the incoming/outgoing wave flowing through the TL.
Along the TL, the voltage $v\left(z,t\right)$ and current $i\left(z,t\right)$
can be written in terms of the superposition of incoming and outgoing
voltage waves $v^{+}\left(z,t\right)$ and $v^{-}\left(z,t\right)$
as $v\left(z,t\right)=v^{+}\left(z,t\right)+v^{-}\left(z,t\right)$
and $i\left(z,t\right)=i^{+}\left(z,t\right)+i^{-}\left(z,t\right)=v^{+}\left(z,t\right)/Z_{0}-v^{-}\left(z,t\right)/Z_{0}$,
where $Z_{0}$ is the characteristic impedance of TL. In turn, the
real voltage waves $v^{\pm}\left(z,t\right)$ can be separated into
the complex wave amplitude $S^{\pm}\left(z,t\right)$ and its complex
conjugate as $v^{\pm}\left(z,t\right)=\sqrt{\frac{Z_{0}}{2}}\left[S^{\pm}\left(z,t\right)+S^{\pm}\left(z,t\right)^{*}\right]$.
We assume that $S^{\pm}\left(z,t\right)=\tilde{S}^{\pm}\left(z,t\right)e^{j\omega_{0}t}$
with a slow envelope $\tilde{S}^{\pm}\left(z,t\right)$ such that
$\frac{\partial S^{\pm}}{\partial t}\approx j\omega_{0}S^{\pm}$.
Correspondingly, the time-averaged incoming and outgoing power with
respect to the period $2\pi/\omega_{0}$, i.e., $P^{\pm}=\left\langle v^{\pm}i^{\pm}\right\rangle $,
are simply $\pm\left|S^{\pm}\right|^{2}$, benefiting from the proper
normalization factor in the definition of wave amplitudes. From now
on, we will use $S^{\pm}\left(t\right)$ to represent the incoming
and outgoing wave amplitude at the ending position $z=0$ of the TL,
where the external coupling capacitor $C_{e}$ is attached. The set
of circuit equations accounting for the coupling between the TL and
the LC resonator are
\begin{align}
i\left(0,t\right) & =C_{e}\frac{d}{dt}\left[v\left(0,t\right)-v\right]\label{eq:eC1}\\
\frac{di\left(0,t\right)}{dt} & =C\frac{d^{2}v}{dt^{2}}+\frac{v}{L}\label{eq:eC2}
\end{align}
where $v$ is the node voltage of the LC resonator. Assuming that
$Z_{0}\sim O\left(z_{0}\right)$ with $z_{0}\equiv\sqrt{L/C}$ and
$\epsilon\rightarrow0$ for the weak coupling, we can use the complex
mode amplitude $\psi$ of the resonator and the input/output wave
amplitude $S^{\pm}$ to reformulate Eqs. (\ref{eq:eC1}) and (\ref{eq:eC2})
as

\begin{align}
S^{-} & \approx S^{+}+j\sqrt{\omega_{0}r}\epsilon\psi\\
\frac{d\psi}{dt} & \approx j\omega_{0}\left(1-\frac{1}{2}\epsilon\right)\psi-\frac{1}{2}\omega_{0}r\varepsilon^{2}\psi+j\sqrt{\omega_{0}r}\epsilon S^{+}
\end{align}
where $r=Z_{0}/z_{0}\sim O\left(1\right)$, and the rotating-wave
approximation enabled by the weak-coupling assumption is employed
in the derivation.

Finally, we study the effect of a small driving on the dynamics of
the LC resonators for two relevant cases. 
We start by considering a LC resonator with
time-dependent capacitance $C\left(t\right)=C\left[1+\delta\left(t\right)\right]$
with $\delta\left(t+2\pi/\Omega\right)=\delta\left(t\right)$. Under
the weak and slow driving assumptions such that $\delta\left(t\right)\rightarrow0$
and $\Omega/\omega_{0}\rightarrow0$, we can rewrite the circuit equation
$v/L+\frac{d}{dt}\left[C\left(t\right)v\right]=0$ using the complex
mode amplitude $\psi\left(t\right)$ as
\begin{align}
\frac{d\psi}{dt} & \approx\left[j\left(1-\frac{1}{2}\delta\right)\omega_{0}-\dot{\delta}\right]\psi.
\end{align}
Therefore, the driving could introduce effective gain/loss to the
system in addition to modifying the resonant frequency.

Next, we consider the case of
two identical LC resonators coupled through a time-modulated capacitance 
$C_k \left(t\right)=C\left[h_0 + h_1 \left(t\right)\right]$, with $h_1 \left(t+2\pi/\Omega\right)=h_1\left(t\right)$. 
Like in the previous case, we resort to the approximations $h_1\left(t\right)\rightarrow0$ and $\Omega/\omega_{0}\rightarrow0$, and weak coupling limit $h_0 \rightarrow 0$, 
which allow us to write the Kirchoff equations as described by 
Eqs. (\ref{eq: circuit1a}) and (\ref{eq: circuit1b}) but replacing 
$\kappa \rightarrow h(t)=h_0 + h_1(t)$.
Introducing the complex mode amplitudes $\psi_1$ and $\psi_2$, 
\begin{eqnarray}
 \frac{d \psi_1}{dt} &\approx&  j\omega_0 \left(1-\tilde{h}(t)\right)  \psi_1 
 +j\omega_0 \tilde{h}(t) \psi_2, \\
  \frac{d \psi_2}{dt} &\approx&  j\omega_0 \tilde{h}(t) \psi_1 +j\omega_0 \left(1-\tilde{h}(t)\right)  \psi_2 ,
\end{eqnarray}
where $\tilde{h}(t)=\frac{h(t)}{2}-j \frac{\dot h(t)}{\omega_0}$.
Like in the previous case of the driven LC resonator, the driving of the capacitance 
introduces effective gain/loss due to the non-zero imaginary part of $\tilde{h}(t)$.
But in contrast, the driving of the coupling capacitance not only
modulates the coupling but also the resonant frequencies of the resonators.

In the weak coupling limit, above mechanisms can be superimposed on
top of each other independently ignoring higher-order effects. For
example, using this rule we can write down directly the CME for the
circuits as shown in Fig. 1(c,d), where we set 
$C_{n}\left(t\right)=C\left[1+\delta_{n}\left(t\right)\right]$, $n=2,3$,
and $C_k \left(t\right)=C \ h(t)=C\left[h_0 + h_1 \left(t\right)\right]$.
For simplicity, we consider $h_0=\kappa$.
Explicitly, using the complex mode amplitudes of each resonator $\left\langle n\right.\left|\psi\right\rangle =\psi_{n}$
and the input/output wave amplitudes $\left\langle n\right.\left|S^{\pm}\right\rangle =S_{n}^{\pm}$,
we have
\begin{align}
\frac{d}{dt}\left|\psi\right\rangle  & =\left[jH_{0}\left(t\right)-\Gamma\right]\left|\psi\right\rangle +jD^{T}\left|S^{+}\right\rangle \\
\left|S^{-}\right\rangle  & =\left|S^{+}\right\rangle +jD\left|\psi\right\rangle 
\end{align}
where \\
 \small{$H_{0}\left(t\right)=\omega_{0}
\begin{bmatrix}
w_1  & \frac{\kappa}{2} & \frac{\kappa}{2}\\
\frac{\kappa}{2} & w_1-\frac{\delta_{2} + h_1}{2}   & \frac{\kappa}{2}+\frac{h_1}{2} \\
\frac{\kappa}{2} & \frac{\kappa}{2}+\frac{h_1}{2} & w_1-\frac{\delta_{3} + h_1}{2}  
\end{bmatrix}$}, \\
with $w_1=1-\kappa-\frac{1}{2}\epsilon$,\\
{\footnotesize$\Gamma=\frac{1}{2}D^{\dagger}D+
\begin{bmatrix}
0 & 0 & 0\\
0 & \dot \delta_2 + \dot h_1 & -\dot h_1\\
0 & -\dot h_1 &  \dot \delta_3 + \dot h_1
\end{bmatrix}$}, \\ 
{\footnotesize$D=\begin{bmatrix}\sqrt{\omega_{0}r}\epsilon & 0 & 0\\
0 & \sqrt{\omega_{0}r}\epsilon & 0\\
0 & 0 & \sqrt{\omega_{0}r}\epsilon
\end{bmatrix}$}. 
\\
Note that the precise form of the CMT used in the main text can
be obtained by simply letting $\left|S^{\pm}\right\rangle \rightarrow\mp j\left|S^{\pm}\right\rangle $,
i.e., a proper redefinition of phase factors.


\begin{thebibliography}{1}
\bibitem{efftemp} The effective temperature is associated with commercially available noise generators with noise power of $-95dBm/Hz$.
\end{thebibliography}
\end{document}